# Defect Localization in Flip-Chip Devices Using Space-Domain Reflectometry and Magnetic Current Imaging

David Noel
Department of Electrical and Computer Engineering, University of Florida

**Abstract**

*Magnetic Field Imaging (MFI) is the newest Fault Isolation/Failure analysis technique to non-destructively and non-invasively localize defects such as electrical shorts and opens in both the die and package levels of Flip-Chips. This is accomplished using Magnetic Current Imaging (MCI) and Space-Domain Reflectometry (SDR) techniques accompanied using a Giant Magneto-Resistor (GMR), which provides detailed spatial optical images at sub-micron resolutions to localize further and identify defects [1]. This paper will demonstrate the use of MCI to locate electrical shorts by imaging the magnetic fields induced by the current-carrying wire bonds using a Superconducting Quantum Interference Device (SQUID) [2], then implement SDR to locate electrical opens using SQUID and GMR to produce detailed optical images of the defect locations. The exact location of the defect can then be localized by using a CAD overlay of the circuit schematic with the optical and MCI images.*

## 1. INTRODUCTION

By Moore's law, which states that the number of transistors on a semiconductor device will double every 18 months, chip designers have adopted very complex methods of implementing said transistors; the latest method being 3-D stacking of devices which increases the number and densities of transistors and other devices within one package, thus increasing system performance in terms of power consumption, speed, and noise reduction. However, this design configuration quickly adds new metal layers and interconnects, posing huge reliability issues as many of these interconnects tend to weaken, migrate, and become disconnected at the package or die levels, creating shorts or opens within the circuitry. This has inherently led to the development of new failure analysis tools which seek to quickly identify, isolate and resolve these issues, with the latest of these tools being Magnetic microscopy, which encompasses MCI (*for localizing electrical shorts and leakages*), SDR (*for locating electrical opens*) and GMR (*for producing sub-micron resolution optical images*).

Magnetic Microscopy is a contactless, non-destructive, and non-invasive technique used in defect localization [3]. Unlike many of the previous tools used in failure analysis, Magnetic microscopy can locate and isolate defects at both the die and package levels with highly sensitive magnetic sensors, which form the core of Magnetic Field Imaging, can be used to identify these defects at both the die and package levels.

### 1.1. Magnetic Field Imaging Principle of Operation

Magnetic Field Imaging (MFI) is based on creating a map of the magnetic fields produced by current-carrying structures. The major advantage of magnetic fields is that they can pass through virtually any material used in the chip fabrication process without being disturbed. These "mapped" magnetic fields are then transformed into a current density image using a Fast Fourier Transform Inversion procedure. This mapped current can then be used to localize short circuits or opens and verify that charges flow in the circuit's areas where expected.

The basic idea of the principle is that when a charge flows through a conductor, it produces a magnetic field that obeys the Biot-Savart law:

$$d\vec{B} = \frac{\mu_0 I}{4\pi} \frac{d\vec{l} \times \vec{r}}{r^2} \quad (1)$$

Where *B* is the induced magnetic field, *I* is the current amplitude, *l* is the current direction, and *r* is the current length. By measuring the *z* component of the magnetic fields produced, the current distribution, *I*, can then be determined; a problem is the Magnetic Inverse Problem, which is solved by the Magnetic Current Imaging (MCI) technique, which is independent of the sensors used. To calculate this current distribution, however, a few geometrical and electrical assumptions must be made, with the most important being that the charges under investigation must all flow in the same *x-y* plane [4].

To drastically reduce the number of possible solutions to the magnetic inverse problem, a simulation model must also be developed (this is built into the MCI system), and the final solution that would create the 3-D current map would be the blending or trade-off between acquisition parameters such as working distance, noise, current amplitude, number of metal lines and the simulation parameters such as simulation time and number of simulations.

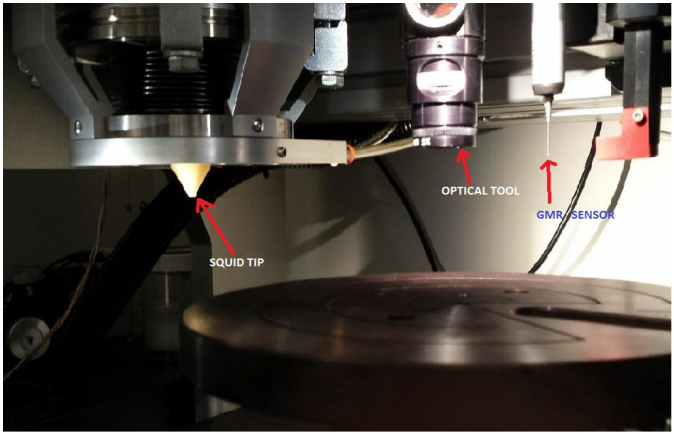

**Figure 1.** Setup of MCI with SQUID

Several sensors obtain the magnetic fields, the most sensitive being SQUID (Superconducting Quantum Interference Device). This magnetic sensor allows the mapping of extremely weak currents (500 nA) at very large working distances from the device under test (DUT) [5]. This is accomplished using superconducting Josephson junctions, which must operate at cryogenic temperatures and be housed in a vacuum with the magnetic sensor tip. The DUT is AC-biased to reduce the effects of noise by using lock-in amplification.

The other sensor is a Giant Magneto-Resistor (GMR) which utilizes magnetic tunnel junctions to provide high-definition optical resolutions. This sensor, however, must be closer to the DUT than the SQUID sensor. The MCI technique is now used in several industrial devices to generate highly detailed bi-directional current maps by quantifying the available magnetic fields above a device without having to de-package that device.

## 2. SPACE DOMAIN REFLECTOMETRY

Space domain reflectometry is part of the MFI analysis, a novel idea currently used in Failure Analysis to detect and localize electrical opens. Using an RF probe, this technique injects a continuous radio frequency (RF) signal (60 – 200) MHz through a defective trace. Since an electrical open defect acts like a boundary where all incident waves are reflected, the RF signal at these locations practically disappears.

The SQUID can then be used to image the RF signals at these locations in 2D, and that image can be overlaid with a CAD circuit image for precise localization of the failure. See Figure 2 for a high-level schematic of the setup required to use SDR with SQUID for locating open defects.

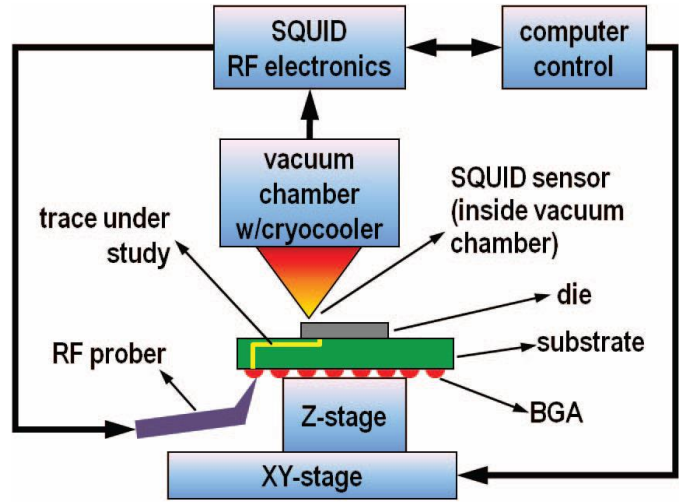

**Figure 2.** Schematic of SDR setup to locate opens using SQUID system[1].

### 2.1 Principle of SDR

An open defect can be modeled as a resistor with resistance **R** in parallel with a capacitor of capacitance **C**, and therefore its total impedance **Z** can be modeled by the equation:

$$Z = \frac{R}{1+(\omega RC)^2} + i\frac{\omega R^2 C}{1+(\omega RC)^2} \qquad (2)$$

Where, $\omega$, is the angular frequency of the RF signal. Depending on the magnitude $\omega RC$, the open can be primarily capacitive or resistive. Typically, the impedance of the open is in the Mega-Ohms (MΩ) range.

When the RF signal is injected into the defective trace, which has a characteristic impedance, $Z_0 \approx 50\Omega$, and encounters an open, all the RF power would be reflected towards the probe because $Z \gg Z_0$. When this happens, the incident and reflected waves superimpose to form a standing wave [6], forming a node at the open defect, as shown in Figure 3.

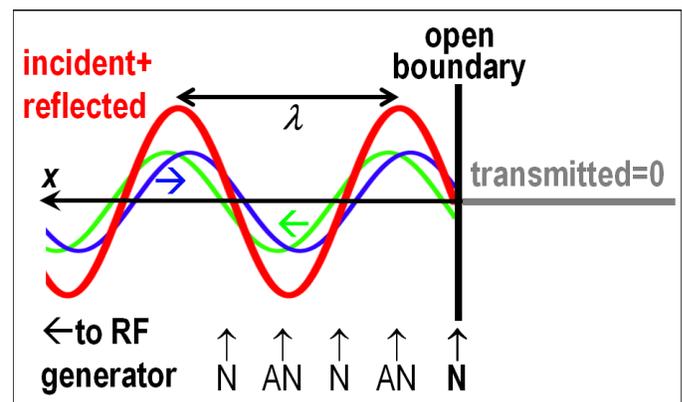

**Figure 3.** Illustration of standing wave (red) created by the superposition of reflected (green) and incident (blue) waves at an open defect[1].

The magnitude of the current, *I*, of the standing wave produced versus its position, *z*, along the defective trace can be described by the equation:

$$I = 2\sqrt{\frac{P}{Z_0}} \sin(\beta z) \approx 2\sqrt{\frac{P}{Z_0}} \beta z \qquad (3)$$

Where P is the probe's incident power and $\beta = \frac{2\pi}{\lambda}$ is the wavenumber. At RF frequencies, since $z \ll \lambda$, implies that the current, *I*, linearly decays (Linear Decay Model) in the locale of the open because $\sin(\beta z) \approx \beta z$. By using SQUID to map the magnetic field of the standing wave at the area of the open, the current profile of the standing wave can be created, and the open defect can be located within the package or die to an accuracy of 30μm.

### 2.2 Sample Experiment to Determine Opens

The sample used in this experiment was a 5.25cm x 4.5cm Land Grid Array (LGA) multi-core flip-chip. One of the data pins on the chip failed continuity tests to the ground during production testing. This failure was verified further by the Quality and Reliability department and was subsequently analyzed using SDR and SQUID to determine the exact location of the failure within the package. See Figure 4 for the location of the failed pin.

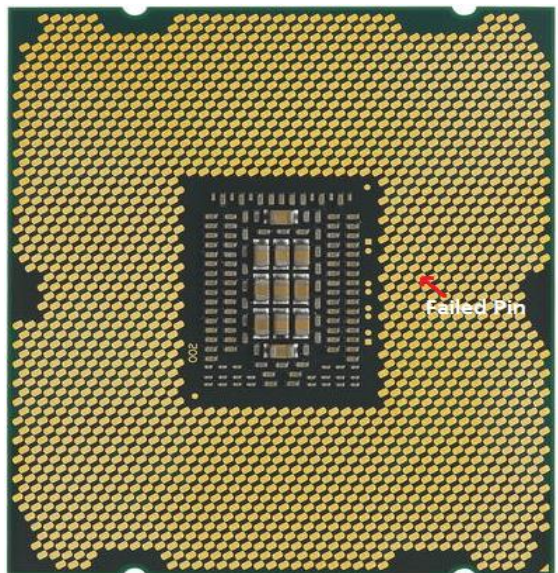

**Figure 4.** Chip showing failed pin.

An SDR probe injected an RF frequency of 60MHz into this pin trace, with the SQUID held at 290μm from the device under test (DUT).

The resulting image in Figure 5 represents the data obtained from the linear decay model showing the standing wave near the open.

From the image, the standing wave decays linearly until it inevitably hits a noise floor which prevents it from decaying to zero linearly, therefore, following the linear portion of the curve, a line (red) was drawn and extrapolated to the zero power location of the curve, which gives the location, z, of the open defect.

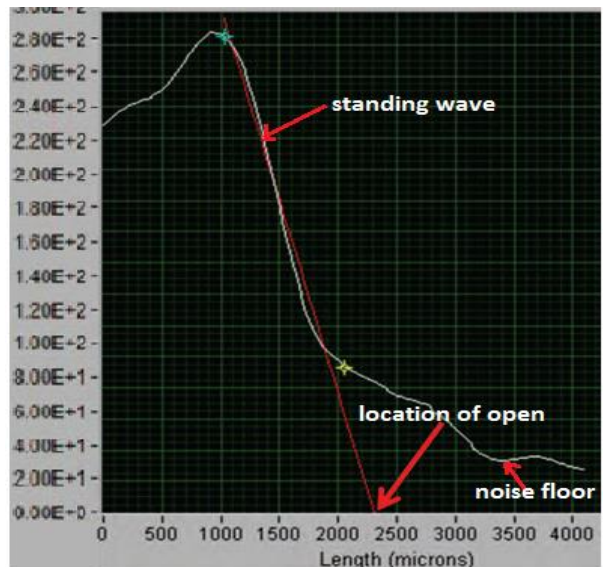

**Figure 5.** Linear decay model showing standing wave and location of the open defect.

After analyzing the linear decay model shown in Figure 5 for the exact location of the open, the SQUID could also show the current map near the open and mark the defect location with a "+" hairline, as shown in Figure 6 below.

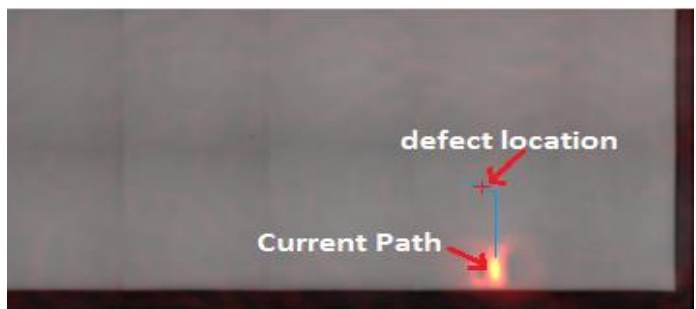

**Figure 6.** SQUID software shows an Image of the current map near an open defect.

The location determined in Figure 5 was projected onto an optical image of the vicinity of the open defect obtained using the GMR sensor shown in Figure 7 below.

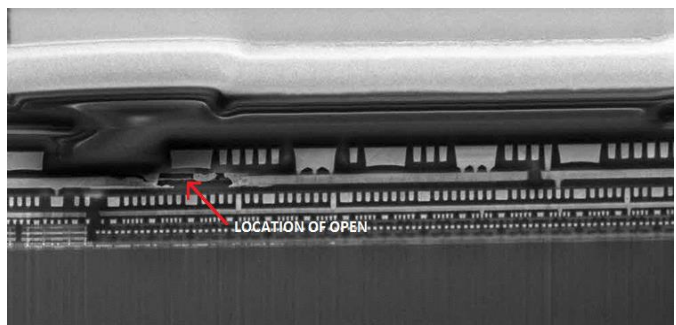

**Figure 7.** Optical image showing open defect location.

Subsequently, after further physical analysis of the above location using a Scanning Electron Microscope (SEM), the defect was found and confirmed to be an electrical open in a metal line, as shown in Figure 8.

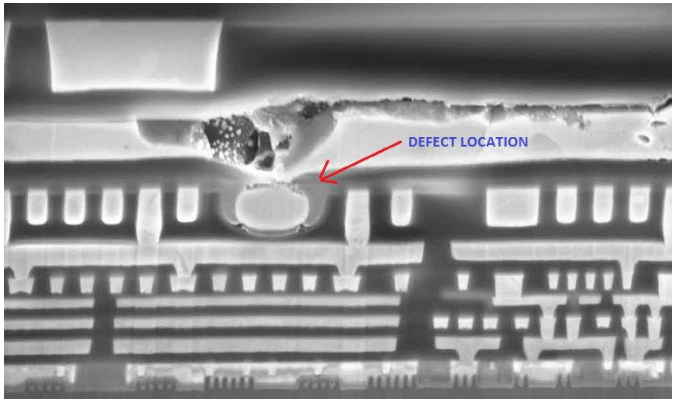

**Figure 8.** SEM image of electrical open defect location.

## 3. ELECTRICAL SHORT LOCALIZATION

In addition to using SDR to locate opens in a package or die, the MFI system utilizes an equally important technique known as MCI (Magnetic Current Imaging) to detect and localize electrical shorts [5]. See Figure 9 for a schematic view of the MCI system.

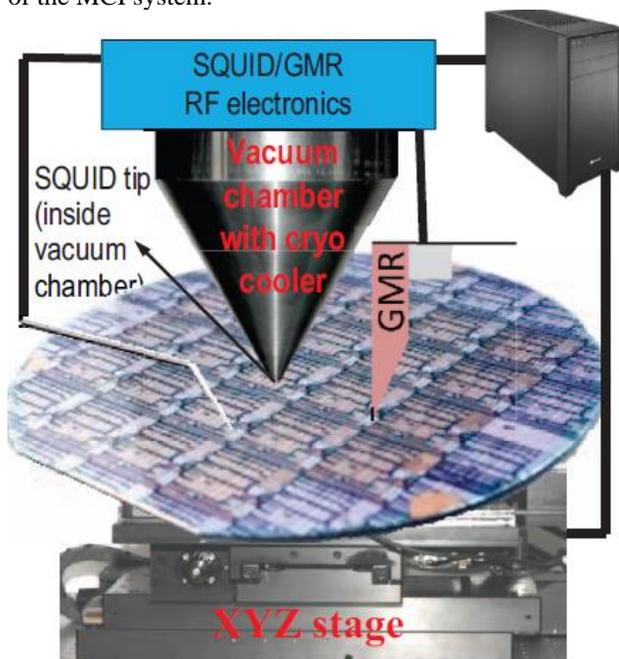

**Figure 9.** Schematic view of MCI system[7].

In this technique, the MCI system uses the SQUID and GMR sensors on the same platform. This combination provides both the current map of the DUT (via SQUID) and its associated high-resolution optical image (via GMR) [4].

### 3.1 Sample Experiment to Determine Electrical Shorts

This sample experiment was done with colleagues in parallel departments to characterize the SQUID system.

The failing unit used was a wafer die that registered a very low resistance (≈1.5KΩ) between two unconnected neighboring pads. Such a low resistance produced a very high current that was abnormal for that piece of circuitry and much larger than the golden reference unit used for comparison.

The pad in question was AC-biased with approximately 9.5 volts at 90 KHz and subjected to a SQUID scan to determine the approximate location of where the short circuit may be occurring. The following figure shows the acquired SQUID image overlaid with the optical image.

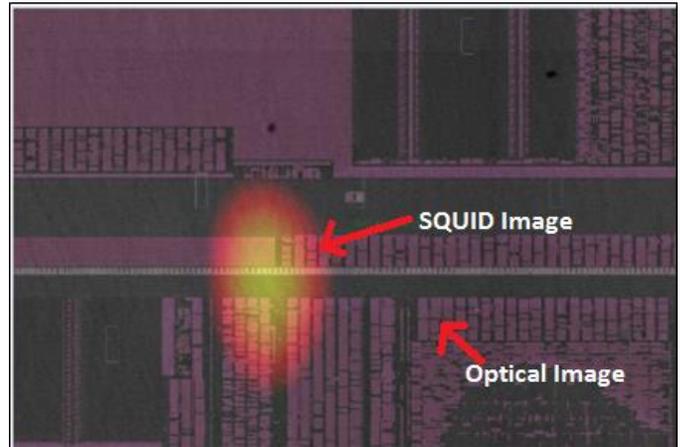

**Figure 10.** SQUID image overlaid with optical image.

As can be seen in Figure 10 above, the area referenced in the lower left side of the image, which contains a loop in the SQUID image, indicates the existence of the short since the current at that point sought a different, less resistive path, than what was seen in the golden(reference) unit.

At this point, the GMR sensor was employed to scan the area where the loop is indicated in Figure 10 to produce a high-resolution optical image [1]. This optical image was then overlaid with that area's SQUID image (which shows current paths). The resulting image is shown in Figure 11.

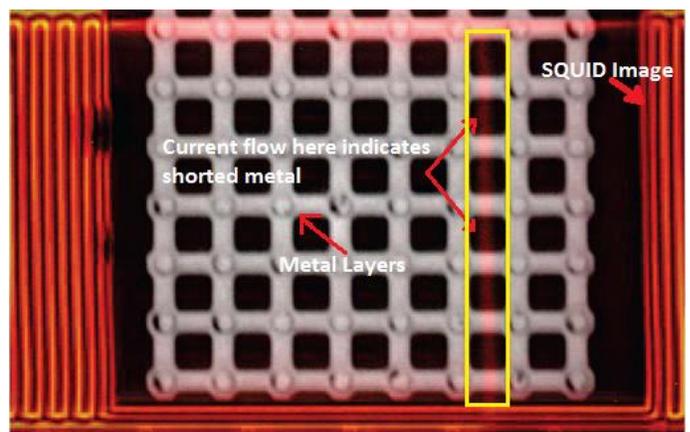

**Figure 11.** SQUID plus GMR Image of the affected area. The region within the yellow block shows shorted area[7].

From Figure 11, the region highlighted within the yellow block shows current flowing across the vertical metal layers, which should not be occurring in this case. This is the area where the short occurs, which was determined using the

SQUID and GMR systems without deconstructing or touching the sample [2].

Now that the defective area was localized using the MFI system, a physical analysis was performed on the sample using very high-resolution microscopes to reveal the structure and nature of the defect only in that region without destroying the sample. The defect can be seen in Figure 12 after microscope analysis.

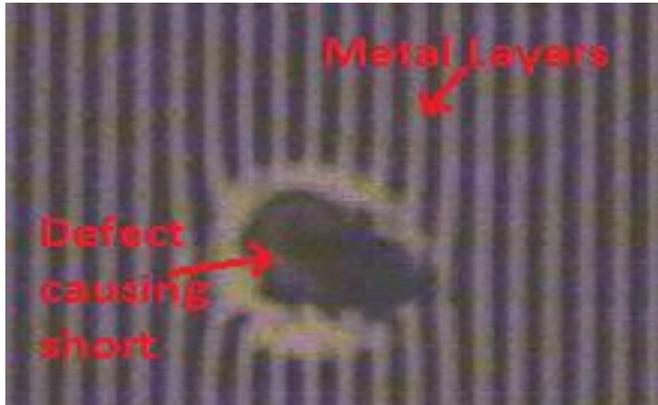

**Figure 12.** Microscope image of the defect causing short[7].

## 4. CONCLUSION

With the emergence of 3-D stacking technologies in chip design and manufacturing, it is becoming increasingly complex to perform fault isolation using traditional methods, such as Infra-Red microscopy alone, due to the increased complexity of the transistor and metal arrangements. Furthermore, many traditional means require contact with the sample, which normally requires that the sample be initially deconstructed, thereby greatly increasing time and cost. With the advent of MFI, dominant defects in 3-D semiconductor packages, such as electrical shorts and opens, can be found expeditiously without first deconstructing the DUT.

The above experiments have shown that fault isolation using MFI is a very suitable technique for isolating difficult-to-find defects in emerging multi-arranged die packages.

ACKNOWLEDGMENT

I want to acknowledge the generous support and access given to me by my manager and mentors at Intel Corporation in the continued research efforts of this project.